\documentclass[12pt,aps,showpacs]{revtex4}%
\usepackage{amsmath}
\usepackage{amsfonts}
\usepackage{amssymb}
\usepackage{graphicx}%
\providecommand{\U}[1]{\protect\rule{.1in}{.1in}}
\begin{document}
\title{Incompressible Stars and Fractional Derivatives}
\author{S. S. Bayin}
\affiliation{Institute of Advanced Mathematics, Middle East\ Technical University, Ankara,
06800, Turkey}
\author{J. P. Krisch}
\affiliation{Department of Physics, University of Michigan, Ann Arbor, MI 48109}
\date{10 March, 2014}

\begin{abstract}
Fractional calculus is an effective tool in incorporating the effects of
non-locality and memory into physical models. \ In this regard, successful
applications exist ranging from signal processing to anomalous diffusion and
quantum mechanics. \ In this paper we investigate the fractional versions of
the stellar structure equations for non radiating spherical objects. \ Using
incompressible fluids as a comparison, we develop models for constant density
Newtonian objects with fractional mass distributions or stress conditions. To
better understand the fractional effects, we discuss effective values for the
density, gravitational field and equation of state. \ The fractional objects
are smaller and less massive than integer models. \ The fractional parameters
are related to a polytropic index for the models considered.

\end{abstract}

\maketitle

\section{Introduction}

Fractional calculus offers a convenient way to introduce memory, non locality
and other fractional effects into physical models that are not covered by
standard non-linear but local models \cite{OS74, RH2000}. Fractional
generalizations of some of the basic differential equations of physics have
led to new understandings of the dynamics underlying macroscopic phenomena in
a wide range of areas \cite{IP99, SKB2002, KST06} like anomalous diffusion,
signal processing and quantum mechanics \cite{NL2000}. Applications to stellar
structure include the work of El-Nabulsi \cite{EN2011}, \ who considered a
fractional equation of state for white dwarf stars. \ Two of the fundamental
tools in building models for compact objects are the equation of hydrostatic
equilibrium (HSE) and the spherical symmetry condition (mass-density
relation). \ In this paper we develop some new models for Newtonian
incompressible, static and spherically symmetric stars by fractionalizing
these two conditions. \ Constant density, incompressible fluids are useful in
modeling because they provide analytic solutions that can be compared to the
.behavior of real physical systems \cite{TS1998, PPF04}. \ Solid planets,
white dwarfs and neutron stars may be approximated by incompressible fluid
models \cite{PPF04,FW05}. Recent work has linked the Navier-Stokes equations
for an incompressible fluid to solutions of the Einstein field equations in
higher dimensions \cite{LS11, BKLS11, HLPTW11}. \ Incompressible fluids have
also been linked to other model fluid descriptions \cite{MN0012}.
\ Fractionalizing the structure equations for a Newtonian constant density
star provides some new stellar models with interesting features not found in
constant density stars described by the integer structure equations. \ 

Because of high densities, some stellar models use general relativity to
incorporate curvature effects. Naturally, the general hydrostatic equilibrium
equation for Newtonian fluids follow from the zero covariant divergence of the
stress energy tensor, $T^{ij}{}_{;\text{ }j}=0.$ \ Fractionalizing partial
derivatives is usually just a replacement of an integer derivative with a
fractional derivative. \ Fractionalizing covariant derivatives in general
relativity is more complex than fractionalizing a partial derivative because
of the contributions of Christoffel connections to the covariant derivative
and to the structure of the tensor definitions. Spherically symmetric
relativistic and non-radiating stellar models usually start with the metric
\cite{ZN71}
\begin{equation}
ds^{2}=-e^{\nu(r)}c^{2}dt^{2}+(1-2Gm(r)/c^{2}r)^{-1}dr^{2}+r^{2}d\Omega^{2},
\end{equation}
and the Einstein field equations are solved for $\nu(r)$ and $m(r)$ for a
given equation of state. The general integer HSE equation is derived and then
the Newtonian limit is taken before fractionalizing. \ We consider three
cases: \ a fractional HSE equation and integer mass relation, an integer HSE
equation and fractional mass relation and both HSE and the mass relation
fractionalized. \ In all three cases, the fractional models can describe
objects that are smaller and less massive than those described by integer
models. \ 

In the next section we briefly review the integer models. \ The fractional
models are developed in the third part of the paper where we also compare the
fractional model details to the standard description of an incompressible
fluid star. \ In Section 4 we develop an effective value for density which
allows a non-infinite effective sound speed to be defined. \ Some possible
reasons for the differences between fractional and integer stars are discussed
in the last part of the paper. We use the Caputo fractional derivative
\cite{RH2011} to develop the models. \ A brief review of the differences
following from using the Caputo, Riemann-Liouville and\ Riesz derivatives is
in the Appendix. \ 

\section{Incompressible Stars}

\subsection{HSE and Mass relation}

The HSE condition assumes a perfect fluid stress energy content for the star
\begin{equation}
T^{ij}=(\rho+\frac{P}{c^{2}})U^{i}U^{j}+Pg^{ij},
\end{equation}
where $P$ and $\rho$ are the pressure and the density distributions,
respectively. $U^{i}$ is the four velocity and $g^{ij}$ is the metric tensor.
Using the zero divergence condition provides the equation
\begin{equation}
(\rho+\frac{P}{c^{2}})_{,j}U^{i}U^{j}+(\rho+\frac{P}{c^{2}})(U_{;j}^{i}%
U^{j}+U^{i}U_{;j}^{j})+P_{,j}g^{ij}=0
\end{equation}
with spacelike $(h_{ik}=g_{ik}+U_{i}U_{k})$ and timelike $(U_{i}U_{k})$
projections, respectively, as $\ $%
\begin{align}
(\rho+\frac{P}{c^{2}})U_{;j}^{k}U^{j}+P,_{j}g^{kj}  &  =0,\\
\rho,_{j}U^{j}+(\rho+\frac{P}{c^{2}})U_{;j}^{j}  &  =0.
\end{align}
\ The field equations for the metric, Eq. (1), provide the relation between
the mass $m(r)$ inside a radius $r,$ and the density as
\begin{equation}
\frac{dm(r)}{dr}=4\pi\rho r^{2}.
\end{equation}

Using this, the relativistic hydrostatic equilibrium equation, Eq.(4), is
written as
\begin{equation}
\frac{dP}{dr}=-\frac{G(\rho+\frac{P}{c^{2}})}{r^{2}}(1-\frac{2Gm}{rc^{2}%
})^{-1}(\frac{4\pi r^{3}P}{c^{2}}+m(r)).
\end{equation}
The vacuum boundary at $r=R$ is defined as $P(R)=0$ and the total mass of the
star is defined as $M=m(R)$. \ The mass and pressure gradient at the star's
center are both zero.
\begin{equation}
m(0)=0\text{ and }\left.  \frac{dP}{dr}\right\vert _{r=0}=0.
\end{equation}
In the classical limit, $P<<\rho c^{2}$ , $2Gm/rc^{2}<<1$ , Eq (7) reduces to
the Newtonian hydrostatic equilibrium equation
\begin{equation}
\frac{\partial P}{\partial r}=-G\frac{m(r)}{r^{2}}\rho,
\end{equation}
Eqs.(6,9) are the basic stellar structure equations for a spherically
symmetric non radiating Newtonian star which are to be solved for a given
equation of state.

\subsection{Comparison Model: \ Incompressible stars}

Incompressible stars have constant density, $\rho=\rho_{c}$. \ This is the
model we use for comparison in fractionalizing the structure equations. For
each of the fractional models considered, we compare the mass, pressure and
radius of the integer star to the fractional star. \ The standard constant
density parameters are%

\begin{align}
m(r)  &  =\frac{4\pi\rho_{c}}{3}r^{3},\\
P(r)  &  =P_{c}-\frac{2\pi G\rho_{c}^{2}}{3}r^{2},
\end{align}

The pressure is zero at the boundary of the star, $r=R,$ and this allows the
central pressure to be written in terms of the constant density and radius as
\begin{equation}
P_{c}=\frac{2\pi G\rho_{c}^{2}}{3}R^{2}%
\end{equation}
The central pressure will take this value for all models. \ The rapidity of
the rise to the zero pressure surface will vary in the fractional models. The
radius of the integer star is \
\begin{equation}
R=\left(  \frac{3P_{c}}{2\pi G\rho_{c}^{2}}\right)  ^{1/2},
\end{equation}
where the mass of the star is%
\begin{equation}
M=4\pi\rho_{o}\int_{0}^{R}r^{2}dr=\frac{4\pi}{3}\left(  \frac{3P_{c}}{2\pi
G}\right)  ^{3/2}\frac{1}{\rho_{c}^{2}}%
\end{equation}
and the $M-R$ relation is given as%
\begin{equation}
M=\frac{2}{G}\left(  \frac{P_{c}}{\rho_{c}}\right)  R.
\end{equation}
The quantity in brackets, $\frac{P_{c}}{\rho_{c}},$ or $M/R$, is a measure of
the compactness of the star as measured by the Einstein red-shift \cite{ST83}.
The larger the ratio, the more mass can be contained within a given radius. \ 

\section{Fractional Stellar Structure Equations}

Fractionalizing the partial derivative in the spherical symmetry mass
relation,\ Eq.(6), and hydrostatic equilibrium, Eq. (9), introduces a unit
inconsistency. \ One way to keep consistent units while fractionalizing is to
express the derivatives in terms of a dimensionless quantity. \ The
derivatives in Eq. (6, 9) can be rewritten using a scale parameter
\begin{equation}
\chi=r/R
\end{equation}
$,$ where $R$ is the radius of the integer star, Eq. (13). \ The rescaled
equations that will be fractionalized are
\begin{equation}
\frac{dm(\chi)}{d\chi}=4\pi\rho R^{3}\chi^{2}.
\end{equation}%
\begin{equation}
\frac{\partial P}{\partial\chi}=-\frac{G}{R}\frac{m(\chi)}{\chi^{2}}\rho.
\end{equation}

The scaled radius of the star, following from $P=0,$ will be denoted by
$\chi_{o}$. \ The three models we consider are a fractional HSE condition with
an integer mass derivative, a fractional mass derivative with an integer HSE
condition and both mass and HSE fractionalized. \ The models we develop will
reproduce the integer Newtonian results as the fractional parameters approach
their integer limit. \ 

\subsection{Model 1: \ Fractional HSE and Integer Mass-Density relation}

Replacing the partial derivative in Eq. (9) with a Caputo derivative (see
Appendix), the fractional generalization of the hydrostatic equilibrium
equation is%
\begin{equation}
\left(  \frac{d^{\alpha}P_{(\alpha)}}{d\chi^{\alpha}}\right)  _{C}=-\frac
{G}{R}\frac{m_{(\alpha)}(\chi)}{\chi^{2}}\rho_{(\alpha)},\text{ }0<\alpha
\leq1.
\end{equation}
The $(\alpha)$ subscript on the mass and stress identify fractional
parameters. The pressure, $P_{(\alpha)},$ the mass, $m_{(\alpha)},$ and the
density, $\rho_{(\alpha)},$ have their usual units. \ Using a dimensionless
coordinate for the fractional derivative allows $G$ to have its usual
Newtonian units. \ The Caputo derivative and its Laplace transform are
defined, respectively, as%
\begin{align}
\left[  \frac{d^{q}f(t)}{dt^{q}}\right]  _{C}  &  =\frac{1}{\Gamma(1-q)}%
\int_{0}^{t}\left(  \frac{df(\tau)}{d\tau}\right)  \frac{d\tau}{(t-\tau)^{q}%
},\text{ }0<q\leq1,\\
\pounds \left\{  \left[  \frac{d^{q}f(t)}{dt^{q}}\right]  _{C}\right\}   &
=s^{q}\widetilde{f}(s)-s^{q-1}f(0),\text{ }0<q\leq1,
\end{align}
where $\widetilde{f}(s)$ is the Laplace transform of $f(t).$ For the first
model we keep the spherical symmetry condition as%
\begin{equation}
\frac{dm_{(\alpha)}(\chi)}{d\chi}=4\pi R^{2}\chi^{2}\rho_{(\alpha)}.
\end{equation}
Assuming constant density, $\rho_{c},$ we write%
\begin{equation}
m_{(\alpha)}(\chi)=4\pi\chi^{3}R^{3}\rho_{c}/3.
\end{equation}
With this mass function, the fractional stress gradiant is
\begin{equation}
\left(  \frac{d^{\alpha}P_{(\alpha)}}{d\chi^{\alpha}}\right)  _{C}=-\frac{4\pi
GR^{2}\rho_{c}^{2}}{3}\chi,\text{ }0<\alpha\leq1.
\end{equation}
\ Taking the Laplace transform of the fractional hydrostatic equilibrium
equation with the boundary condition (or performing the fractional integral),
the pressure is%
\begin{equation}
P_{(\alpha)}(\chi)=-\frac{4\pi GR^{2}\rho_{c}^{2}}{3\Gamma(\alpha+2)}%
\chi^{\alpha+1}+P_{c},
\end{equation}

\ Finding the radius from the surface condition, $P_{(\alpha)}(\chi
_{o(\alpha)})=0,$ we write%
\begin{equation}
\chi_{o(\alpha)}^{\alpha+1}=\frac{\Gamma(\alpha+2)}{2},
\end{equation}
which indicates that the fractional object is smaller than its integer
counterpart, $\chi_{o(\alpha)}^{\alpha+1}<1.$ \ The total mass, $M_{\alpha},$
is%
\begin{align}
M_{(\alpha)}  &  =m_{(\alpha)}(\chi_{o(\alpha)}),\\
M_{(\alpha)}  &  =M(\frac{\Gamma(\alpha+2)}{2})^{3/(1+\alpha)}.
\end{align}
This fractional star has a smaller mass and a smaller radius than the integer
star, $M_{(\alpha)}<M.$ \ 

Examining Eq.(24) one might assume, with no explicit dependence on the
fractional parameter, that the fractional and integer stress gradients were
the same. \ However, even with no explicit fractional dependence, the integer
and fractional derivative operations are very different and result in stresses
with a strong fractional dependence. \ For this model, the low alpha
fractional pressure ($P_{(\alpha)}(\chi)-P_{c}$ ) is larger and increases more
rapidly than the integer pressure as $\chi$ decreases toward the origin. \ In
Part 4 of the paper, we will show that Model 1 objects with their constant
density, can be described as an n=0 polytrope. \ Normally, the $n=0$
polytropes are models for planet-like objects. \ 

\subsection{Model 2: \ Integer HSE and Fractional Mass-Density relation}

We now consider the case where the scaled stellar structure equations are
given as%
\begin{align}
\frac{dP_{(\beta)}}{d\chi}  &  =-G\frac{m_{(\beta)}(\chi)}{R\chi^{2}}%
\rho_{(\beta)},\\
\left(  \frac{d^{\beta}m_{(\beta)}}{d\chi^{\beta}}\right)  _{C}  &  =4\pi
\chi^{2}R^{3}\rho_{(\beta)},\text{ }0<\beta\leq1.
\end{align}
Assuming $\rho_{(\beta)}=\rho_{c},$ we generate the following stellar model:%
\begin{align}
P_{(\beta)}(\chi)  &  =-8\pi GR^{2}\rho_{c}^{2}\left[  \frac{\Gamma(\beta
+1)}{\Gamma(\beta+3)\Gamma(\beta+2)}\right]  \chi^{1+\beta}+P_{c},\\
m_{(\beta)}(\chi)  &  =8\pi R^{3}\rho_{c}\frac{\chi^{2+\beta}}{\Gamma
(\beta+3)},\\
\chi_{o(\beta)}^{1+\beta}  &  =\frac{\Gamma(\beta+3)\Gamma(\beta+2)}%
{12\Gamma(\beta+1)}.
\end{align}
The mass of the star is
\begin{equation}
M_{(\beta)}=M(\frac{\Gamma(\beta+3)\Gamma(\beta+2)}{12\Gamma(\beta
+1)})^{1/(1+\beta)}\frac{\Gamma(\beta+2)}{2\Gamma(\beta+1)}.
\end{equation}

As in the previous model, this fractional star is smaller and has less mass
than the integer star. \ For this case, with the integer Newtonian HSE
condition, the stress gradiants vary with the fractional parameter because of
the fractional mass-radius relation in this model. \ \ The stresses,
$P_{(\beta)}(\chi)-P_{c}$ , are smaller for this model than for Model 1 and
show a less rapid decrease toward the star's center. \ As in Model 1, the
Model 2 stresses are larger than the integer stress. \ In the 4th part of the
paper, an effective density can be defined for this model. \ \ 

\subsection{Model 3: \ Models with both $0<\alpha\leq1$ and $0<\beta\leq1$}

For the most general case with $\rho_{(\alpha,\beta)}=\rho_{c}$, the stellar
structure equations are both fractional and are given as%

\begin{align}
\ \left(  \frac{d^{\alpha}P_{(\alpha,\beta)}}{d\chi^{\alpha}}\right)  _{C}  &
=-G\frac{m_{(\alpha,\beta)}(\chi)}{R\chi^{2}}\rho_{c}\\
\left(  \frac{d^{\beta}m_{(\alpha,\beta)}}{d\chi^{\beta}}\right)  _{C}  &
=4\pi\chi^{2}R^{3}\rho_{c},
\end{align}

with solution
\begin{align}
P_{(\alpha,\beta)}(\chi)  &  =-\frac{8\pi G_{\ }R^{2}\rho_{c}^{2}\Gamma
(\beta+1)}{\Gamma(\beta+3)\Gamma(\alpha+\beta+1)}\chi^{\alpha+\beta}+P_{c},\\
m_{(\alpha,\beta)}(\chi)  &  =\frac{8\pi\rho_{c}R^{3}}{\Gamma(\beta+3)}%
\chi^{\beta+2},\\
\chi_{o(\alpha,\beta)}  &  =\left[  \frac{\Gamma(\beta+3)\Gamma(\alpha
+\beta+1)}{12\Gamma(\beta+1)\ }\right]  ^{1/(\alpha+\beta)},\\
M_{(\alpha,\beta)}  &  =\frac{6M}{\Gamma(\beta+3)}\left[  \frac{\Gamma
(\beta+3)\Gamma(\alpha+\beta+1)}{12\Gamma(\beta+1)}\right]  ^{\beta
+2/(\alpha+\beta)},\text{ \ }0<\alpha\leq1,\text{ }0<\beta\leq1.
\end{align}

For $\alpha$ or $\beta$ integer, the previous two models are recovered. \ In
the next section we compare the size and mass of all three models. \ 

\subsection{Model Comparison}

Models 1 and 2 differ in the size of the stress and stress gradiants for the
same $\chi$ and fractional parameter, with the fractional HSE models having
the larger values. \ A comparison of the three fractional star radii and mass
to those of the integer star are shown in Figures (1, 2) for the case of equal
fractional index, $\alpha=\beta.$ Figure 1 shows the scaled radius versus the
fractional index. \ All three fractional models are smaller than their integer
counterpart. \ For the same $\alpha,$ a fractional hydrostatic equilibrium
condition coupled to an integer mass relation will produce a larger star than
the fractional star with an integer hydrostatic equilibrium condition.%
\begin{figure}
[ptb]
\begin{center}
\includegraphics[
trim=0.029017in 0.130419in -0.029018in -0.130419in,
height=4.2823in,
width=4.2823in
]%
{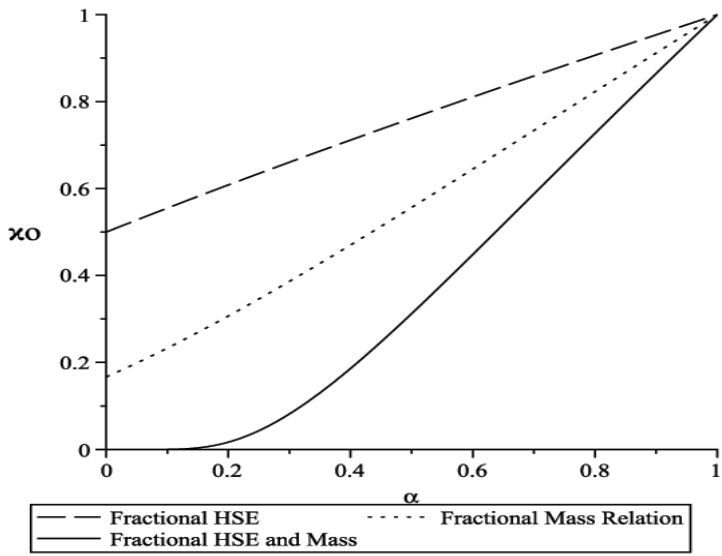}%
\caption{Scaled surface radius $\chi_{o}=R_{frac}/R$ vs.$\alpha$ for the three
models considered.}%
\end{center}
\end{figure}

$\ $

Figure 2 shows the ratio of the fractional mass models to the integer models .
\ All three fractional models have a smaller mass and radius than their
integer counterpart; \ for $\alpha=\beta$, the fractional HSE condition (Model
1) star has the largest mass, radius and stress of the three models fractional
models considered.%

\begin{figure}
[ptb]
\begin{center}
\includegraphics[
trim=0.028906in 0.144983in -0.028906in -0.144983in,
height=4.2823in,
width=4.2823in
]%
{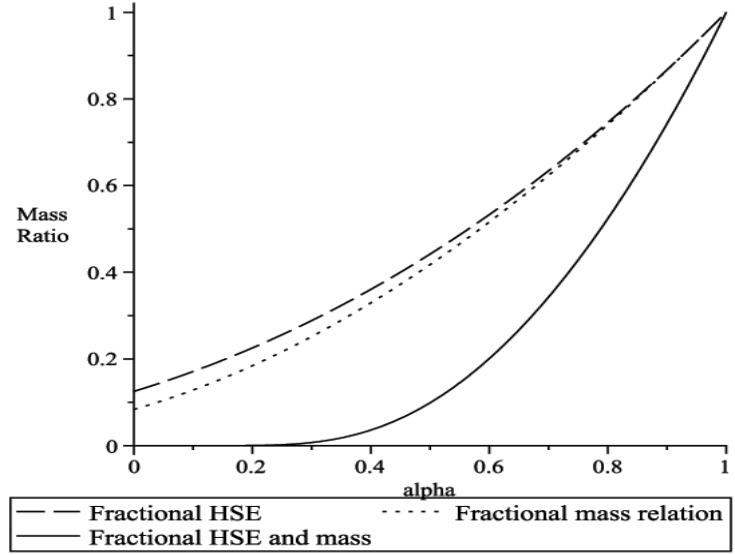}%
\caption{Fractional Mass/Iinteger Mass for the three models considered}%
\end{center}
\end{figure}

Another way of comparing the three models is with an average density. \ The
model results are all scaled in \ terms of a comparison integer incompressible
star. \ Using the general third model, an average density can be defined%
\[
\frac{\rho_{av(\alpha,\beta)}}{\rho_{c}}=\frac{M_{(\alpha,\beta)}}{4\pi
\chi_{o(\alpha,\beta)}^{3}/3}=\frac{6}{\Gamma(\beta+3)}\left[  \frac
{\Gamma(\beta+3)\Gamma(\alpha+\beta+1)}{12\Gamma(\beta+1)}\right]
^{\beta-1/(\alpha+\beta)},\
\]
where the constant density for the integer stars defined by Eq. (10) has been
used as a comparison. \ For very fractional stars (low values of $\alpha$ and
$\beta$), the ratio can be very large. For example, for $\alpha=\beta=0.1,$
the ratio is about 6000 while for higher values of the fractional parameters,
the ratio can be slightly larger than 1. \ The comparison object can be an
ordinary star or a model astrophysical compact object like a planet, a white
dwarf, a neutron star or a hybrid compact object \cite{FS2005,VBLS09}. \ One
should note that the average density for the model 1 stars ($\beta=1),$ is the
same as the integer value so this definition of average density will not
explain any Model 1 differences. \ \ In the next section we discuss systems
equivalent to the fractional models. \ Using an effective equation of state,
we will show that the Model 1 stars are effective n=0 polytropes while Model 2
stars have a polytropic index covering a range of fractional values. \ 

\section{Equivalent Systems}

Comparing fractional mass and radii to integer values is one way to show the
differences between fractional and integer incompressible stars. A way to
understand the differences is to introduce effective values of the fluid
parameters, the values the parameters would have if the structure equations
were not fractionalized. \ We begin by defining an effective density and then
look at a range of effective gravitational accelerations.

\subsection{Effective Density}

An effective density can be found starting with Eq. (30) for constant density
and taking another fractional derivative\
\begin{equation}
\left(  \frac{d^{1-\beta}}{d\chi^{1-\beta}}\right)  _{C}\left[  \left(
\frac{d^{\beta}m}{d\chi^{\beta}}\right)  _{C}\right]  =4\pi R^{3}\left(
\frac{d^{1-\beta}}{d\chi^{1-\beta}}\right)  _{C}\left[  \chi^{2}\rho
_{c}\right]
\end{equation}
and using the relation \cite{SB2006}
\begin{equation}
\left(  \frac{d^{1-\beta}}{d\chi^{1-\beta}}\right)  _{C}\left[  \left(
\frac{d^{\beta}m}{d\chi^{\beta}}\right)  _{C}\right]  =\frac{dm}{d\chi}%
-\frac{\left.  \left(  \frac{d^{\beta}m}{d\chi^{\beta}}\right)  _{C}%
\right\vert _{\chi=0}}{\Gamma(\beta)\chi^{1-\beta}},\text{ }0<\beta<1.
\end{equation}
Since for $\rho=\rho_{c},$
\begin{equation}
\left.  \left(  \frac{d^{\beta}m(\chi)}{d\chi^{\beta}}\right)  _{C}\right\vert
_{\chi=0}=0,
\end{equation}
we can relate the integer mass derivative to the fractional result%
\begin{align}
\frac{dm}{d\chi}  &  =4\pi R^{3}\rho_{c}\left(  \frac{d^{1-\beta}}%
{d\chi^{1-\beta}}\right)  _{C}\left[  \chi^{2}\right]  =4\pi R^{3}\rho
_{c}\frac{2\chi^{1+\beta}}{\Gamma(2+\beta)}=4\pi Rr^{2}\left[  \frac{2\rho
_{c}\chi^{\beta-1}}{\Gamma\left(  \beta+2\right)  }\right]  ,\\
\frac{dm}{dr}  &  =4\pi r^{2}\rho_{c}\frac{2\chi^{\beta-1}}{\Gamma(2+\beta)}.
\end{align}
This allows us to define an effective density in terms of the ordinary
spherical symmetry condition as%
\begin{equation}
\frac{dm}{dr}=4\pi r^{2}\rho_{eff.}(\chi),
\end{equation}
where%
\begin{equation}
\rho_{eff.}=\left[  \frac{2\rho_{c}\chi^{\beta-1}}{\Gamma\left(
\beta+2\right)  }\right]  .
\end{equation}
This is the equivalent mass density that one would find in terms of the
ordinary, non-fractional, spherical symmetry condition. \ It diverges at the
origin but the mass is well behaved. \ This definition will only apply to
Model 2 objects.

\subsection{Effective Equation of State}

Using the fractional stress and the effective density, an effective equation
of state can be written down and used to place limits on the range of the
fractional parameters. \ Using the pressure relation, Eq. (37), the general
fractional radius, Eq. (39) and the effective density we can write an equation
of state
\begin{equation}
P_{(\alpha,\beta)}(\chi)-P_{c}=-P_{c}\frac{12\Gamma(\beta+1)}{\Gamma
(\beta+3)\Gamma(\alpha+\beta+1)}(\frac{\Gamma\left(  \beta+2\right)  }%
{2})^{\frac{\alpha+\beta}{\beta-1}}(\frac{\rho_{eff.}}{\rho_{c}}%
)^{\frac{\alpha+\beta}{\beta-1}}.
\end{equation}
This is in a "polytropic" form relative to pressure gauged to the central
value.
\begin{equation}
P_{(\alpha,\beta)}(\chi)-P_{c}=K\rho_{eff.}^{(n+1)/n}%
\end{equation}
\ with polytropic index
\begin{equation}
n=(1-\beta)/(1+\alpha)
\end{equation}
\ The models we have described are for an incompressible fluid which is
usually an $n=0$ standard polytrope. \ Here the gauge polytrope, $n=0,$
corresponds to a fractional HSE condition (Model 1) with the expected constant
density across the incompressible fluid and an integer mass relation,
$\beta=1,$ $0<\alpha\leq1$. \ The second model, a fractional mass relation,
corresponds to a general polytrope with index\ $n=(1-\beta)/2.$ \ The $n=0$
polytrope can be used to describe spherical planet-like objects while the
higher values can describe stars, their compactness and their modes of core
heat transport. \ White dwarf models have indices greater than 1. \ Neutron
stars have polytropic indices ranging from about 1/2 to 1. \ Here, the largest
values of $n$ corresponds to very low values of $(\alpha,\beta)$ and will
describe fractional polytropes. \ 

A simple calculation of the speed of sound, $V_{s}$, would give $V_{s}%
->\infty$ for an incompressible fluid. \ An effective density associated with
a fractional object allows a finite effective speed of sound from this
equation of state
\begin{equation}
V_{s}^{2}=\frac{\partial P_{(\alpha,\beta)}(\chi)}{\partial\rho_{eff.}}%
=\frac{P_{c}}{\rho_{c}}\frac{6\Gamma(\beta+1)(\alpha+\beta)\Gamma\left(
\beta+2\right)  }{\Gamma(\beta+3)\Gamma(\alpha+\beta+1)(1-\beta)}\chi
^{\alpha+1}\leq c^{2}.
\end{equation}
The acoustic speed cannot exceed light speed and this provides, at the
surface, an inequality on the $P_{c}/\rho_{o}$ ratio for Models 2 and 3 \
\begin{equation}
\frac{P_{c}}{\rho_{c}c^{2}}\leq\frac{2(1-\beta)}{\Gamma(\beta+2)(\alpha
+\beta)}(\frac{12\Gamma(\beta+1)}{\Gamma(\beta+3)\Gamma(\alpha+\beta
+1)})^{(1-\beta)/(\alpha+\beta)}.
\end{equation}

In considering Newtonian stars, the HSE condition follows from the TOV
equation with the condition $P<<\rho c^{2}.$ Graph 3 shows the maximum
$P_{c}/\rho_{c}c^{2}$ ratio\ versus $\alpha$ for three choices of beta.
\ $P_{c}$/$\rho_{c}$ are input model parameters, with $P_{c}$ set by its
integer value. \ From Eq. (15), the integer ratio is
\[
\frac{P_{c}}{\rho_{c}}=\frac{2\pi G\rho_{c}R^{2}}{3}=\frac{GM}{2R}.
\]
The ratio has a very small upper limit for $\beta$ and $\alpha$ close to their
integer value. \ This does not preclude choosing smaller values of
$(\alpha,\beta)$ in building fractional models so long as the actual value of
$P_{c}/\rho_{c}$ is small and under its limit. \ The Newtonian fractional
models will certainly be useful in describing small deviations from the
integer values. \ Olson \cite{TSO2000} has discussed equations of state for
maximally incompressible neutron stars in the low temperature limit using a
stricter $V_{s}/c$ limit based on stellar stability. \ A general relativistic
treatment could provide a higher limit parameter range for these fractional
spherical objects.%
\begin{figure}
[ptb]
\begin{center}
\includegraphics[
trim=-0.058035in 0.130545in 0.058035in -0.130545in,
height=4.2823in,
width=4.2823in
]%
{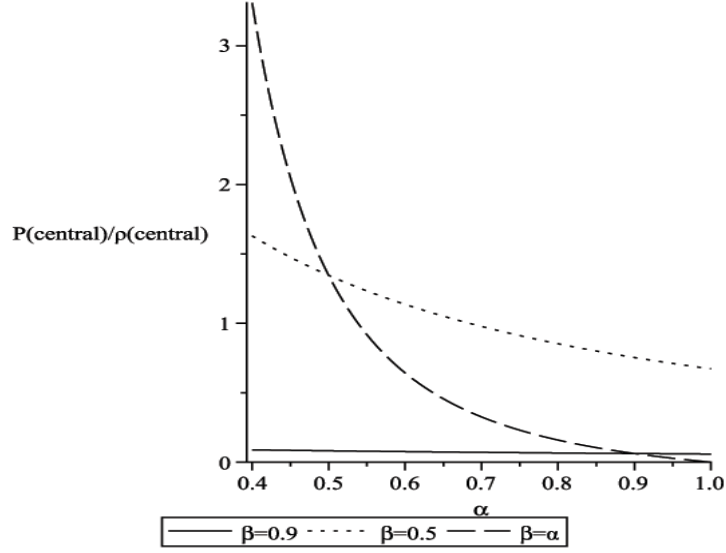}%
\caption{Central Pressure/Central Density vs. $\alpha$ for three choices of
$\beta$}%
\end{center}
\end{figure}
\ 

\subsection{Effective g-field}

An effective gravitational acceleration can also be defined. \ We use two
methods: (1) solving a fractional Poisson equation and (2) from a relation
between the fractional and integer derivatives using the effective density.
\ The two methods produce the same result when $\rho_{eff}$ $=\rho_{c}.$

\subsubsection{Fractional Poisson\ Equation}

The Poisson equation for the gravitational acceleration is%

\begin{equation}
\frac{dg}{dr}=-4\pi G\rho.
\end{equation}
The fractional equation with constant density is written as%
\begin{equation}
\frac{d^{\gamma}g}{d\chi^{\gamma}}=-4\pi GR\rho_{c}%
\end{equation}
with solution%
\[
g_{(\gamma)}=\frac{4\pi GR\rho_{o}\chi^{\gamma}}{\Gamma(1+\gamma)}.
\]
Dividing by the integer value, $g,$ gives
\begin{equation}
4\pi Gr\rho_{c}\frac{g_{(\gamma)}}{g}=\frac{\chi^{\gamma-1}}{\Gamma(1+\gamma
)}.
\end{equation}

\subsubsection{The HSE Condition}

Fractionalizing the mass, density affects the hydrostatic equilibrium
condition and this can be used to get an effective value for the gravitational
acceleration. \ We first write the ordinary hydrostatic equilibrium equation,
Eq. (9), for $\alpha=1$ as%
\begin{equation}
\frac{dP}{dr}=g(r)\rho.
\end{equation}
Following the same procedure that was used to find an effective density, we
relate the regular stress derivative to the fractional Caputo derivative with
the relation
\begin{equation}
\frac{dP}{d\chi}=\left(  \frac{d^{1-\alpha}}{d\chi^{1-\alpha}}\right)
_{C}\left[  \left(  \frac{d^{\alpha}P}{d\chi^{\alpha}}\right)  _{C}\right]
+\frac{\left.  \left(  \frac{d^{\alpha}P}{d\chi^{\alpha}}\right)
_{C}\right\vert _{\chi=0}}{\Gamma(\alpha)\chi^{1-\alpha}},\text{ }0<\alpha
\leq1.
\end{equation}
For constant density, with the mass boundary condition $m(0)=0$, Eq. (8), the
second term vanishes. \ Substituting$\ $the fractional pressure derivative
from Eq. (32) \ along with the fractional mass relation from Eq.(41) we have
into the above equation we find%
\begin{align}
\frac{dP_{(\alpha,\beta)}(\chi)}{d\chi}  &  =\left(  \frac{d^{1-\alpha}}%
{d\chi^{1-\alpha}}\right)  _{C}(-\frac{8\pi G\rho_{c}^{2}R^{2}}{\Gamma
(\beta+3)}\chi^{\beta}),\\
\frac{dP_{(\alpha,\beta)}}{d\chi}  &  =-\frac{8\pi G\rho_{c}^{2}R^{2}%
\Gamma(\beta+1)}{\Gamma(\beta+3)\Gamma(\beta+\alpha)}\chi^{\alpha+\beta-1}.
\end{align}
Rewriting in terms of a regular coordinate derivative this becomes \
\begin{equation}
\frac{dP_{(\alpha,\beta)}(\chi)}{dr}=-\frac{8\pi G\rho_{c}^{2}R\Gamma
(\beta+1)}{\Gamma(\beta+3)\Gamma(\beta+\alpha)}\chi^{\alpha+\beta-1}.
\end{equation}

Using the effective density, one could write%
\begin{equation}
\frac{dP_{(\alpha,\beta)}(\chi)}{dr}=-g_{eff}(\alpha,\beta)\rho_{eff}(\beta),
\end{equation}
giving an effective acceleration%
\begin{equation}
g_{eff(\alpha,\beta)}(\chi)=4\pi G\rho_{c}R\frac{\Gamma(\beta+1)\Gamma
(\beta+2)}{\Gamma(\beta+3)\Gamma(\beta+\alpha)}\chi^{\alpha}.
\end{equation}
Taking the ratio to the regular value of $g_{(1,1)}(\chi)=4\pi G\rho_{c}R\chi$
gives%
\begin{equation}
\frac{g_{eff(\alpha,\beta)}(\chi)}{g_{(1,1)}(\chi)}=\frac{3\Gamma
(\beta+1)\Gamma(\beta+2)}{\Gamma(\beta+3)\Gamma(\beta+\alpha)}\chi^{\alpha-1}.
\end{equation}
Note that for $\beta=1$%
\begin{equation}
\frac{g_{eff(\alpha,1)}(\chi)}{g_{(1,1)}(\chi)}=\frac{\chi^{\alpha-1}}%
{\Gamma(1+\alpha)},
\end{equation}
which is the value obtained by directly fractionalizing the Poisson equation
with $\gamma=\alpha$. \ For $r<R$, effective g-field , $g_{(\alpha,1)}$ has a
stronger power law dependence than the integer g-value. \ \
\[
g_{eff(\alpha,1)}(r)=-\frac{Gm(r)}{R^{\alpha-1}\Gamma(1+\alpha)r^{3-\alpha}}%
\]%
\begin{equation}
\frac{1}{r^{2}}\rightarrow\frac{1}{r^{3-\alpha}}.
\end{equation}

The effective gravitational acceleration, along with the effective density,
provides some insight into the origins of the Model 1 and Model 2
fractional/integer differences. \ Eq. (25) gives the stress for the fractional
Model 1 star. \ If an ordinary derivative wrt $\chi$ is taken, as expected it
is
\[
\frac{dP_{(\alpha)}}{dr}=-\rho_{c}g_{eff(\alpha,1)}(r).
\]
This is the standard form for the stress gradients needed to support the
stellar material. \ The larger Model 1 stresses and smaller more massive stars
compared to integer models can be motivated by the higher effective g-values.
\ \ A similar calculation using\ Eq. (31) for the Model 2 pressure gives a
similar equation but involving the effective density for Model 2, Eq. (47),%
\[
\frac{dP_{(b)}}{dr}=-\rho_{eff}g_{eff(\alpha,1)}(r)
\]

While the effective density for the Model 2 stars is larger than the constant
density for Model 1 stars, the Model 2 effective gravity is smaller over the
low beta ranges than in Model 1. \ The larger effective gravitational
acceleration produces the larger stresses and stress gradients for the Model 1
stars. \ 

\section{Conclusions}

We have presented some model objects obeying the fractional stellar structure
equations. \ All of the three models considered produce fractional objects
that are smaller and less massive than an integer object. \ The stresses for
the three models can be written in terms of the central pressure and radius
as
\begin{align}
\text{Fractional HSE}\text{: }  &  \ P_{(\alpha)}(\chi)=P_{c}[1-(\frac{\chi
}{\chi_{o(\alpha)}})^{1+\alpha}]\\
\text{Fractional Mass}\text{: }  &  \ P_{(\beta)}(\chi)=P_{c}[1-(\frac{\chi
}{\chi_{o(\beta)}})^{1+\beta}]\\
\text{Both Fractional}\text{: }  &  \ P_{(\alpha,\beta)}(\chi)=P_{c}%
[1-(\frac{\chi}{\chi_{o(\alpha,\beta)}})^{\alpha+\beta}]
\end{align}

The discussion of differences between the fractional stars and their integer
counterparts has focused on comparisons and equivalent systems. \ The
effective density for \ the fractional mass Model 2 stars is larger than for
Model 1 (constant $\rho_{c}$ ) while the effective gravitational accelerations
for fractional HSE Model 1 stars can be much larger than for Model 2 objects.

Beyond these comparisons, is there an underlying property of the fractional
derivative that is driving comparitive differences between the fractional and
integer stars? \ Fractional derivatives are non local in that the derivative
involves an integral over a spatial or time region. \ For static stars, there
is no time, hence memory effects do not enter. \ However, while the fractional
derivatives of mass and stress are given at a particular $\chi=r/R,$ \ they do
spatially sample the entire star. \ For Models 2 and 3 ($\beta\neq1),$the
average density is an indicator of the sampling differences with very
fractional stars having much larger average densities than integer stars. \ A
possible explanation is that the actual mass distibutions of fractional Model
2 and 3 objects are very different than in integer object. \ The fractional
objects could simply be more closely packed than in integer objects. \ A piece
of evidence that argue against the more compact mass distribution explanation
is the ratio of total mass to surface radius, Eq. (15). \ This is less than
one for Model 2 stars and, if regarded an an indicator of compactness,
indicates that the fractional models are less compact than integer models.
\ For example, the ratio for Sirius B is about 150 times larger than our sun.
\cite{ST83}. \ However, using M/R as an indicator of compactness comes from
red-shift arguments which have not been developed for these simple
non-radiating fractional models. \ 

The effective density and sound speed cannot be defined for the gauge
polytrope, Model 1, which maintains its constant density. \ Given the
difference in the objects usually modeled by gauge and non-gauge polytropic
indices, a direct comparison between Model 1 and the other models may be an
unphysical comparison. A Model 1 comparison could be to planets which are
smaller and less massive than their integer counterparts for a constant
density. \ A constant density planet is a very simple model and the fractional
sampling could be used to model actual differences in structure. \ 

The models presented in this paper used the fractional Caputo derivative.
There are other fractional derivatives, each extending the fractional model
assumptions. \ One possibility is the fractional Riesz derivative. \ Models
developed using this derivative have an effective gravitational constant,
$G_{(\alpha)}$ and a gravitational field very similar to the effective field
considered in the previous section \ An effective gravitational constant also
can appear in theories of dimension $D$ \cite{KT1994, TP2010}. \ For example,
in $D$-dimensions, Gauss's law for spherically symmetric mass distributions
can be written
\begin{equation}
g_{D}(r)=-G_{D}\frac{m(r)}{r^{D-1}},
\end{equation}
with $G_{D}$ the gravitational constant in $D$ space. This also suggests that
for incompressible stars the fractional models are related to incompressible
fluids in dimension $D=4-\alpha$. \ \ Another possibility is the
Riemann-Liouville derivative. \ Its use requires the introduction of
fractional boundary conditions $P^{(\alpha-1)}(0)$ and $m^{(\alpha-1)}(0)$, a
different extension of the Caputo models. \ Both possibilities need further
investigation. \ 

\section{Appendix}

Three of the commonly used fractional derivatives are the Caputo derivative,
the Riemann-Liouville derivative and the Riesz derivative. \ The Caputo
derivative, used in the models constructed in this paper, and was developed by
modifying the Riemann-Liouville derivative. In this Appendix, we briefly
describe the three derivatives, beginning with the Caputo derivative.

\subsection{Caputo Fractional Derivative}

In the 1960's Caputo introduced a new definition of the fractional derivative
\cite{OS74, IP99, RH2000, SB2006,KST06, RH2011}%
\begin{equation}
\left[  \frac{d^{q}f(t)}{dt^{q}}\right]  _{C}=\frac{1}{\Gamma(1-q)}\int
_{0}^{t}\left(  \frac{df(\tau)}{d\tau}\right)  \frac{d\tau}{(t-\tau)^{q}%
},\text{ }0<q<1,
\end{equation}
which was used by him to model dissipation effects in linear
viscosity.\bigskip\ The two derivatives are related by%
\begin{equation}
\left[  \frac{d^{q}f(t)}{dt^{q}}\right]  _{C}=\left[  \frac{d^{q}f(t)}{dt^{q}%
}\right]  _{R-L}-\frac{t^{-q}f(0)}{\Gamma(1-q)},\text{ }0<q<1.
\end{equation}

Laplace transforms of the Riemann-Liouville and the Caputo derivative are
given as \cite{NL2000}
\begin{align}
\pounds \left\{  _{0}^{R-L}\mathbf{D}_{t}^{q}f(t)\right\}   &  =s^{q}%
\widetilde{f}(s)-\sum_{k=0}^{n-1}s^{k}\left.  \left(  _{0}^{R-L}\mathbf{D}%
_{t}^{q-k-1}f(t)\right)  \right\vert _{t=0},\text{ }n-1<q\leq n,\\
\pounds \left\{  _{0}^{C}\mathbf{D}_{t}^{q}f(t)\right\}   &  =s^{q}%
\widetilde{f}(s)-\sum_{k=0}^{n-1}s^{q-k-1}\left.  \frac{d^{k}f(t)}{dt^{k}%
}\right\vert _{t=0},\text{ }n-1<q\leq n,
\end{align}
where $_{a}\mathbf{D}_{t}^{q}f(t)\equiv\dfrac{d^{q}f}{[d(t-a)]^{q}}.$

\subsection{\textbf{Riemann-Liouville Definition of Differintegral:}}

The basic definition of fractional derivative and integral, that is,
differintegral, is the Riemann-Liouville (R-L) definition:

For $\ q<0,$ the R-L fractional integral is evaluated by using the formula
\begin{equation}
\text{\ }\left[  \frac{d^{q}f}{[d(t-a)]^{q}}\right]  =\frac{1}{\Gamma(-q)}%
\int_{a}^{t}[t-t^{\prime}]^{-q-1}f(t^{\prime})dt^{\prime},\text{
\ \ \ \ }\ q<0.
\end{equation}

For fractional derivatives, $q\geq0,$ the above integral is divergent, hence
the R-L formula is modified as \cite{SB2006}
\begin{equation}
\left[  \frac{d^{q}f}{[d(t-a)]^{q}}\right]  =\frac{d^{n}}{dt^{n}}\left[
\frac{1}{\Gamma(n-q)}\int_{a}^{t}[t-t^{\prime}]^{-(q-n)-1}f(t^{\prime
})dt^{\prime}\right]  ,\text{\ \ \ \ }q\geq0,\text{ }n>q,
\end{equation}
where the integer $n$ must be chosen as the smallest integer
satisfying\ $(q-n)<0$.

For $0<q<1$ and $a=0,$ the Riemann-Liouville fractional derivative becomes
\begin{equation}
\left[  \frac{d^{q}f(t)}{dt^{q}}\right]  _{R-L}=\frac{1}{\Gamma(1-q)}\frac
{d}{dx}\int_{0}^{t}\frac{f(t^{\prime})d\tau}{(t-t^{\prime})^{q}},\text{
}0<q<1.
\end{equation}

This provides another approach to fractional structure. The Caputo and the
Riemann-Liouville derivatives are related by%
\begin{equation}
\left[  \frac{d^{q}f(t)}{dt^{q}}\right]  _{C}=\left[  \frac{d^{q}f(t)}{dt^{q}%
}\right]  _{R-L}-\frac{t^{-q}f(0)}{\Gamma(1-q)},\text{ }0<q<1,
\end{equation}
The boundary conditions on the mass and pressure are
\begin{equation}
m(0)=0,\text{ }P(0)=P_{c},
\end{equation}
Using these the two different fractional derivatives can be related and we
have%
\begin{align}
\left[  \frac{d^{\beta}m(r)}{dr^{\beta}}\right]  _{R-L}  &  =\left[
\frac{d^{\beta}m(r)}{dr^{\beta}}\right]  _{C},\text{ }0<\alpha<1,\\
\left[  \frac{d^{\alpha}P(r)}{dr^{\alpha}}\right]  _{R-L}  &  =\left[
\frac{d^{\alpha}P(r)}{dr^{\alpha}}\right]  _{C}-\frac{r^{-\alpha}P_{c}}%
{\Gamma(1-\alpha)},\text{ }0<\alpha<1.
\end{align}
Due to the fact that the Laplace transform of the Riemann-Liouville derivative
is given as%
\begin{equation}
\pounds \left\{  \frac{d^{q}f(t)}{dt^{q}}\right\}  =s^{q}\text{ }\widetilde
{f}(s)-f^{(q-1)}(0),\text{ }0<q<1,
\end{equation}
solving the fractional stellar structure equations in terms of the
Riemann-Liouville definition:%
\begin{align}
\left(  \frac{d^{\alpha}P}{dr^{\alpha}}\right)  _{R-L}  &  =-G_{\alpha}%
\frac{m(r)}{r^{2}}\rho,\\
\left(  \frac{d^{\beta}m}{dr^{\beta}}\right)  _{R-L}  &  =4\pi r^{2}\rho
\end{align}
demands using the boundary conditions in terms of fractional derivatives:%
\begin{equation}
P^{(\alpha-1)}(0)\text{ and }m^{(\alpha-1)}(0),
\end{equation}

\subsection{\textbf{Riesz Derivative}}

The Riesz derivative is defined with respect to its Fourier transform
\cite{NL2000}
\begin{equation}
\mathcal{F}\left\{  \mathbf{R}_{t}^{q}f(t)\right\}  =-\left\vert
\omega\right\vert ^{q}g(\omega),\text{ }0<q<2,
\end{equation}
as
\begin{equation}
\mathbf{R}_{t}^{q}f(t)=-\frac{1}{2\pi}\int_{-\infty}^{+\infty}\left\vert
\omega\right\vert ^{q}g(\omega)e^{i\omega t}d\omega,
\end{equation}
where $g(\omega)$ is the Fourier transform of $f(t)$. Note that
\begin{equation}
\mathbf{R}_{t}^{2}f(t)=\frac{d^{2}}{dt^{2}}f(t).
\end{equation}

The Riesz derivative provides another approach to fractional stellar
structure. We first write the stellar structure equations as%
\begin{align}
\overrightarrow{\nabla}P(\overrightarrow{r})  &  =-\rho\overrightarrow{\nabla
}\phi(\overrightarrow{r}),\\
\nabla^{2}\phi(\overrightarrow{r})  &  =4\pi G\rho(\overrightarrow{r}),
\end{align}
where $\phi(\overrightarrow{r})$ is the gravitational potential and
fractionalize the gravitational field equation Eq. (85). We now use the
fractional generalization of the three dimensional Laplacian in terms of the
Riesz derivative as \cite{NL2000}%
\begin{equation}
\Delta^{\alpha/2}\phi(\overrightarrow{r})=-\frac{1}{\left(  2\pi\right)  ^{3}%
}\int_{-\infty}^{\infty}d^{3}\overrightarrow{k}\text{ }\widetilde{\phi
}(\overrightarrow{k})\left\vert \overrightarrow{k}\right\vert ^{\alpha
}e^{i\overrightarrow{k}\text{\textperiodcentered}\overrightarrow{r}},\text{
}1<\alpha\leq2,
\end{equation}
where $\widetilde{\phi}(\overrightarrow{k})$ is the Fourier transform of
$\phi(\overrightarrow{r}):$%
\begin{align}
\widetilde{\phi}(\overrightarrow{k})  &  =\int_{-\infty}^{\infty}%
d^{3}\overrightarrow{r}\text{ }\phi(\overrightarrow{r})e^{-i\overrightarrow
{k}\text{\textperiodcentered}\overrightarrow{r}},\\
\phi(\overrightarrow{r})  &  =\frac{1}{\left(  2\pi\right)  ^{3}}\int
_{-\infty}^{\infty}d^{3}\overrightarrow{k}\text{ }\widetilde{\phi
}(\overrightarrow{k})e^{i\overrightarrow{k}\text{\textperiodcentered
}\overrightarrow{r}}.
\end{align}
In other words, the Fourier transform of the fractional Laplacian is%
\begin{equation}
\mathcal{F}\{\Delta^{\alpha/2}\phi(\overrightarrow{r})\}=-\widetilde{\phi
}(\overrightarrow{k})\left\vert \overrightarrow{k}\right\vert ^{\alpha}.
\end{equation}
We now consider the fractional generalization of the stellar structure
equations as%
\begin{align}
\overrightarrow{\nabla}P(\overrightarrow{r})  &  =-\rho\overrightarrow{\nabla
}\phi(\overrightarrow{r}),\\
\Delta^{\alpha/2}\phi(\overrightarrow{r})  &  =4\pi G\rho(\overrightarrow{r}).
\end{align}
Taking the Fourier transform of the fractional gravitational field equation,
we write the solution as
\begin{equation}
\phi(\overrightarrow{r})=-\frac{4\pi}{\left(  2\pi\right)  ^{3}}G_{\alpha}%
\int_{-\infty}^{\infty}d^{3}\overrightarrow{k}\text{ }\frac{\widetilde{\rho
}(\overrightarrow{k})}{\left\vert \overrightarrow{k}\right\vert ^{\alpha}%
}e^{i\overrightarrow{k}\text{\textperiodcentered}\overrightarrow{r}},
\end{equation}
where $\widetilde{\rho}(\overrightarrow{k})$ is the Fourier \ transform of the
density distribution. We can also write the above equation as%
\begin{equation}
\phi(\overrightarrow{r})=-\frac{4\pi G_{\alpha}}{\left(  2\pi\right)  ^{3}%
}\int_{-\infty}^{\infty}d^{3}\overrightarrow{r}^{\prime}\text{ }\int_{-\infty
}^{\infty}d^{3}\overrightarrow{k}\text{ }\frac{\ e^{i\overrightarrow
{k}\text{\textperiodcentered}(\overrightarrow{r}-\overrightarrow{r}^{\prime}%
)}}{\left\vert \overrightarrow{k}\right\vert ^{\alpha}}\rho(\overrightarrow
{r}^{\prime})\ ,
\end{equation}
or as
\begin{equation}
\phi(\overrightarrow{r})=-\frac{4\pi G_{\alpha}}{\left(  2\pi\right)  ^{3}%
}\int_{-\infty}^{\infty}d^{3}\overrightarrow{r}^{\prime}\text{ }%
I(\overrightarrow{r}-\overrightarrow{r}^{\prime})\rho(\overrightarrow
{r}^{\prime}),
\end{equation}
where%
\begin{equation}
I(\overrightarrow{r}-\overrightarrow{r}^{\prime})=\int_{-\infty}^{\infty}%
d^{3}\overrightarrow{k}\text{ }\frac{\ e^{i\overrightarrow{k}%
\text{\textperiodcentered}(\overrightarrow{r}-\overrightarrow{r}^{\prime})}%
}{\left\vert \overrightarrow{k}\right\vert ^{\alpha}}.
\end{equation}
Using the substitution $k^{\prime}=k\left\vert \overrightarrow{r}%
-\overrightarrow{r}^{\prime}\right\vert ,$ we evaluate the angular part of the
above integral to write \
\begin{equation}
I(\overrightarrow{r}-\overrightarrow{r}^{\prime})=\frac{\left(  2\pi/i\right)
}{\left\vert \overrightarrow{r}-\overrightarrow{r}^{\prime}\right\vert
^{3-\alpha}}\left[  \int_{-\infty}^{\infty}dk^{\prime}\text{ }\frac
{\ k^{\prime}e^{ik^{\prime}}}{\left\vert k^{\prime}\right\vert ^{\alpha}%
}\right]  .
\end{equation}
Thus,%
\begin{equation}
\phi(\overrightarrow{r})=-\frac{4\pi}{\left(  2\pi\right)  ^{3}}\frac{2\pi}%
{i}G_{\alpha}\int_{-\infty}^{\infty}d^{3}\overrightarrow{r}^{\prime}\text{
}\left[  \int_{-\infty}^{\infty}dk^{\prime}\text{ }\frac{\ k^{\prime
}e^{ik^{\prime}}}{\left\vert k^{\prime}\right\vert ^{\alpha}}\right]
\frac{\rho(\overrightarrow{r}^{\prime})}{\left\vert \overrightarrow
{r}-\overrightarrow{r}^{\prime}\right\vert ^{3-\alpha}}\ .
\end{equation}
When $\alpha=2,$ the $k^{\prime}$ integral can be evaluated as a Cauchy
principal value integral \cite{SB2006} as $\pi i,$ thus yielding the Newtonian
potential:%
\begin{equation}
\phi(\overrightarrow{r})=-G\int_{-\infty}^{\infty}d^{3}\overrightarrow
{r}^{\prime}\frac{\rho(\overrightarrow{r}^{\prime})}{\left\vert
\overrightarrow{r}-\overrightarrow{r}^{\prime}\right\vert }.
\end{equation}
For $1<\alpha<2$ we evaluate the integral as%
\begin{align}
\int_{-\infty}^{\infty}\frac{ke^{ik}}{\left\vert k\right\vert ^{\alpha}}dk  &
=\int_{-\infty}^{0}\frac{ke^{ik}}{\left\vert k\right\vert ^{\alpha}}%
dk+\int_{0}^{\infty}\frac{ke^{ik}}{k^{\alpha}}dk\\
&  =2i\int_{0}^{\infty}k^{2-\alpha}\frac{\sin k}{k}dk\\
&  =2i\ast2^{(1-a)}\sqrt{\pi}\Gamma(3/2-\alpha/2)/\Gamma(\alpha/2).
\end{align}
Thus,
\begin{equation}
\phi(\overrightarrow{r})=-2^{(2-\alpha)}\frac{\Gamma(3/2-\alpha/2)}{\sqrt{\pi
}\Gamma(\alpha/2)}G_{\alpha}\int_{-\infty}^{\infty}d^{3}\overrightarrow
{r}^{\prime}\frac{\rho(\overrightarrow{r}^{\prime})}{\left\vert
\overrightarrow{r}-\overrightarrow{r}^{\prime}\right\vert ^{3-\alpha}}.
\end{equation}
Refining the gravitational constant $G_{\alpha}$ as%
\begin{equation}
\overline{\text{ }G}_{\alpha}=2^{(2-\alpha)}\frac{\Gamma(3/2-\alpha/2)}%
{\sqrt{\pi}\Gamma(\alpha/2)}G_{\alpha}%
\end{equation}
and dropping bar we finally write%
\begin{equation}
\phi(\overrightarrow{r})=-\ G_{\alpha}\int_{-\infty}^{\infty}d^{3}%
\overrightarrow{r}^{\prime}\frac{\rho(\overrightarrow{r}^{\prime})}{\left\vert
\overrightarrow{r}-\overrightarrow{r}^{\prime}\right\vert ^{3-\alpha}}.
\end{equation}


\section{References}

\begin{thebibliography}{99}                                                                                               %


\bibitem {OS74}K. B. Oldham and J. Spanier, \textit{The Fractional Calculus},
Dover (1974).

\bibitem {RH2000}R. Hilfer (ed.), \textit{Fractional Calculus: Applications in
Physics,} World Scientific ( 2000).

\bibitem {SKB2002}Igor M. Sokolov, Joseph Klafter and Alexander Blumen,
Physics Today, November (2002).

\bibitem {KST06}A. A. Kilbas, H. M.\ Srivastava and J. J.\ Trujillo,
\textit{Theory and Applications of Fractional Differential Equations},
Elsevier (2006).

\bibitem {IP99}I. Podlubny, \textit{Fractional Differential Equations},
Academic Press ( 1999).

\bibitem {NL2000}N. Laskin, Phys. Rev. \textbf{E62} (2000) 3135.

\bibitem {EN2011}El-Nabusi Ahmad Rami, Applied Mathematics and Computation
\textbf{218} (2011), 2837. \ The Fractional white dwarf hydrodynamical
nonlinear differential equation and emergence of quark stars. \ 

\bibitem {TS1998}Keisuke Taniguchi and Masaru Shibata, Phys. Rev.
\textbf{D58},084012 (1998). Gravitational radiation from co-rotating binary
neutron stars of incompressible fluid in the first post-Newtonian
approximation of general relativity.

\bibitem {PPF04}P. P. Fiziev, arXiv:astro-ph/0409458 Novel Geometrical Models
of Relativistic Stars. II. Incompressible Stars and Heavy Black Dwarfs

\bibitem {FW05}F. Weber, Prog.Part.Nucl.Phys. \textbf{54},193 (2005). Strange
quark matter and compact stars. \ 

\bibitem {BKLS11}Irene Bredberg, Cynthia Keeler, Vyacheslav Lysov, Andrew
Strominger, arXiv.1101.2451 [hep-th], From Navier-Stokes to Einstein

\bibitem {LS11}Vyasheslav Lysov and Andres Strominger, arXiv.1104.5502
[hep-th], From Petrov-Einstein to Navier-Stokes

\bibitem {HLPTW11}Tai-Zhuo Huang, Yi Ling, Wen-Jian Pan, Yu Tian, Xiao-Ning
Wu,JHEP 1110:079,2011 arXiv:1107.1464 \ From Petrov-Einstein to Navier-Stokes
in\ Spatially Curved Spacetime

\bibitem {MN0012}Yoshinori Matsuo, Makoto Natsuume, Masahiro Ohta, Takashi
Okamura, arXiv:1206.6924 [hep-th] The Incompressible Rindler fluid versus the
Schwarzschild-AdS fluid

\bibitem {ZN71}Ya. B. Zeldovich and I. D. Novikov, Relativistic Astrophysics
Vol1 Stars and Relativity, Chicago Press 1971.

\bibitem {RH2011}Richard Herrmann, \textit{Fractional Calculus: \ An
Introduction for Physicists}, World Scientific (2011), pg.42.

\bibitem {ST83}Stuart L.\ Shapiro and\ Saul A. Teukolsky, \textit{Black Holes,
White Dwarfs and\ Neutron Stars: \ The physics of compact objects}, John Wiley
and\ Sons (1983), pg. 71

\bibitem {FS2005}F. Sandin, Eur.\ Phys. J \textbf{C40} (2005) 15.
\textit{Compact stars in the standard model - and beyond}

\bibitem {VBLS09}Matt Visser, Carlos Barcelo,\ Stefano Liberati, Sebastiano
Sonego, arXiv:0902.0346. \textit{Small, dark, and heavy:But is it a black
hole?}

\bibitem {SB2006}Selcuk Bayin, \textit{Mathematical Models in Science
and\ Engineering}, John Wiley \& Sons (2006).

\bibitem {TSO2000}Timothy S. Olson, Phys.Rev.C63,015802 (2000). \ Maximally
incompressible neutron star matter.

\bibitem {KT1994}Edward\ W. Kolb and Michael S. Turner, \textit{The Early
Universe}, Addison-Wesley Publishing Company (1994), pg. 470.

\bibitem {TP2010}T. Padmanabhan, \textit{Gravitation: \ Foundations and
Frontiers}, Cambridge University Press (2010), pg. 647
\end{thebibliography}
\end{document}